\RequirePackage{amsmath}
\documentclass[runningheads]{llncs}

\usepackage[english]{babel}
\usepackage[T1]{fontenc}
\usepackage[utf8]{inputenc}
\usepackage{amsmath}
\usepackage{amsfonts}
\usepackage{multirow}
\usepackage{graphicx}
\usepackage{array}
\usepackage[colorlinks=true, allcolors=blue]{hyperref}

\providecommand{\keywords}[1]
{
  \small	
  \textbf{\textit{Keywords: }} #1
}
\begin{document}
\title{Parameter-efficient Fine-tuning for improved Convolutional Baseline for Brain Tumor Segmentation in Sub-Saharan Africa Adult Glioma Dataset}
\titlerunning{Parameter-efficient Fine-tuning for BraTS Africa}
\author{Bijay Adhikari \inst{*, 1, 2} \and
Pratibha Kulung\inst{3} \and
Jakesh Bohaju\inst{4} \and
Laxmi Kanta Poudel\inst{5} \and
Confidence Raymond\inst{6,7,8}   \and
Dong Zhang\inst{6,9} \and
Udunna C Anazodo\inst{6,7,8,10,11,12} \and
Bishesh Khanal\inst{13} \and
Mahesh Shakya\inst{13}}
\authorrunning{Adhikari, Kulung, Bohaju, Poudel, Shakya et al.}
\institute{
Nepal Research and Collaboration Center (NRCC), Nepal \and
 Birendra Multiple Campus, Tribhuvan University, Nepal \and
 Institute of Engineering, Purbanchal Campus, Tribhuvan University, Nepal \and Bhaktapur Multiple Campus, Tribhuvan University, Nepal \and
 Gandaki College of Engineering and Science, Pokhara University, Nepal \and
 Multimodal Imaging of Neurodegenerative Diseases (MiND) Lab, Department of Neurology and Neurosurgery, McGill University, Montreal, QC, Canada \and
 Lawson Health Research Institute, London, Ontario, Canada \and
 Medical Artificial Intelligence Laboratory (MAI Lab), Lagos, Nigeria \and
 Department of Electrical and Computer Engineering, University of British Columbia, Vancouver, Canada \and
 Montreal Neurological Institute, McGill University, Montréal, Canada \and
 Department of Medicine, University of Cape Town, South Africa \and
 Department of Clinical Radiation Oncology, University of Cape Town, South Africa \and
 Nepal Applied Mathematics and Informatics Institute for research (NAAMII), Nepal \newline
\email{\{bjayadikari.ba, pratibha.kulu63, jakesh.bohaju, justkantapoudel\}@gmail.com,}
\email{craymon8@uwo.ca, donzhang@ece.ubc.ca, udunna.anazodo@mcgill.ca,}
\email{\{bishesh.khanal, mahesh.shakya\}@naamii.org.np}}
\maketitle
\begin{abstract}
Automating brain tumor segmentation using deep learning methods is an ongoing challenge in medical imaging. Multiple lingering issues exist including domain-shift and applications in low-resource settings which brings a unique set of challenges including scarcity of data. As a step towards solving these specific problems, we propose Convolutional adapter-inspired Parameter-efficient Fine-tuning (PEFT) of MedNeXt architecture. To validate our idea, we show our method performs comparable to full fine-tuning with the added benefit of reduced training compute using BraTS-2021 as pre-training dataset and BraTS-Africa as the fine-tuning dataset. BraTS-Africa consists of a small dataset (60 train / 35 validation) from the Sub-Saharan African population with marked shift in the MRI quality compared to BraTS-2021 (1251 train samples). We first show that models trained on BraTS-2021 dataset do not generalize well to BraTS-Africa as shown by 20\% reduction in mean dice on BraTS-Africa validation samples. Then, we show that PEFT can leverage both the BraTS-2021 and BraTS-Africa dataset to obtain mean dice of 0.8 compared to 0.72 when trained only on BraTS-Africa. Finally, We show that PEFT (0.80 mean dice) results in comparable performance to full fine-tuning (0.77 mean dice) which may show PEFT to be better on average but the boxplots show that full finetuning results is much lesser variance in performance. Nevertheless, on disaggregation of the dice metrics, we find that the model has tendency to oversegment as shown by high specificity (0.99) compared to relatively low sensitivity(0.75). The source code is available at \href{https://github.com/CAMERA-MRI/SPARK2024/tree/main/PEFT_MedNeXt}{https://github.com/CAMERA-MRI/SPARK2024/tree/main/PEFT\_MedNeXt}   
% 1. segmentation challenging on sub-saharan africa dataset
% 2. Implementatiopn of MedNeXt with convolution adapter
% 3. evaluation Dice Similarity Coefficient ET-0.73, TC-0.79, WT-0.88

\end{abstract}
\section{Introduction}
Brain Tumor poses a significant global health challenge with glioma being most prevalent, malignant and having poor prognosis resulting in 80\% of glioma patients succumbing to death within two years of diagnosis\cite{poon2020longer}. Additionally, low-to-middle income countries (LMICs), particularly Sub-Saharan Africa (SSA), face a higher disease burden due to limited access to imaging devices and specialists, as well as the usual delayed disease presentation\cite{anazodo2023framework,aderinto2023navigating}. In fact, glioma death rate has risen in SSA, unlike in high-income countries where they continue to decrease\cite{patel2019global}. Brain Tumor Segmentation is an essential component in disease management useful in quantification, radiation therapy planning, and treatment evaluation requiring manual delineation by radiologists\cite{mostafa2023brain}. However, increasing incidence and death rates have resulted in increased workload necessitating automated segmentation methods.

This has spurred the introduction of automated methods including pre-deep learning-based semi- and fully automatic methods such as thresholding\cite{otsu1975threshold}, watershed \cite{fayzi2023introducing}, active contours\cite{kass1988snakes}, atlas-based\cite{bach2015atlas} segmentation approaches were proposed. However, these approaches often face challenges in dealing with anatomical variations and complexities, especially pathological ones,  present in medical images. The limitations of these approaches have led to the exploration of deep learning based solution which have demonstrated promising results in tumor segmentation tasks \cite{abdusalomov2023brain}.

Deep-learning methods, especially Convolutional Neural Network(CNN) based architectures, have emerged as powerful tools for computer vision tasks, including dense prediction tasks such as segmentation. Among them, U-Net\cite{ronneberger2015u} architecture is the most common and widely used baseline model in medical image segmentation due to its parameter efficiency, numerous available implementations, and reasonable performance across varied datasets. It is an encoder-decoder structure with skip connections that efficiently combines spatial information\cite{zargari2023brain} from different scales. However, understanding long-distance spatial relation\cite{kumar2023residual,sahli2023skin}, token-flatten and scale-sensitivity\cite{he2023u} remain challenges for U-Net architecture in brain tumor segmentation. On the other hand, Transformer-based architectures\cite{hatamizadeh2021swin} can leverage long-range dependencies but are highly over-parameterized and require large GPU Memory for training three-dimensional (3D) Volumes.

Towards achieving the right balance, MedNeXt\cite{roy2023mednext}, is a modern convolutional architecture designed for medical image segmentation, inspired by Transformer-based models but tailored for limited annotated datasets. MedNeXt features a fully ConvNeXt 3D Encoder-Decoder Network with Residual ConvNeXt blocks to maintain semantic richness across scales. These blocks mirror Transformer structures with Depthwise Convolution, Expansion Layer, and Compression Layer, facilitating width and receptive field scaling during up- and downsampling\cite{roy2023mednext,maani2024enhancing}. MedNeXt balances computational efficiency and segmentation accuracy, making it ideal for resource-constrained environments like SSA and other LMICs. It addresses the limitations of U-Net by enhancing performance on limited datasets through iterative kernel size increase and compound scaling, providing a robust solution for medical image segmentation\cite{roy2023mednext}.

In this paper, we apply a lightweight architecture MedNeXT \cite{roy2023mednext}, on the BraTS Challenge on Sub-Saharan-Africa adult glioma Dataset\cite{adewole2023brain} to automatically segment glioma into its sub-regions. We aim to demonstrate that models pre-trained on a larger dataset (BraTS-2021) may generalize poorly to smaller and local datasets (BraTS-Africa), underscoring the need for domain adaptation. To address this, we introduce a Parameter-Efficient Fine-Tuning (PEFT) approach, utilizing convolutional adapter-inspired modules integrated with the MedNeXT architecture. The PEFT method achieved comparable, or in some cases, slightly superior performance to full fine-tuned model on the BraTS-Africa dataset, while offering improved computational efficiency.

\section{Related Works}
\subsection{Transfer Learning and Fine-tuning}
Fine-tuning\cite{talukder2023efficient} is a common approach in medical imaging, enabling models to perform well even with limited task-specific datasets. This process involves initializing a model with weights learned from a pre-trained model, and then further training it on a new dataset with task-specific labels. However, full fine-tuning which involves updating all parameters of a pre-trained model, presents some limitations. While this method achieves high performance on downstream tasks, it is parameter-inefficient. The entire model's parameters are updated, leading to increased computational costs and a higher risk of overfitting, particularly when dealing with small datasets. \cite{houlsby2019parameter} proposed a more efficient approach by introducing adapter modules. Their approach demonstrated that fine-tuning all parameters for each task is not always necessary and can result in excessive computational demands. Instead, by training only a small set of additional parameters specific, their method achieved performance comparable to full fine-tuning while significantly reducing the computational burden.

\subsection{Parameter Efficient Fine-tuning}
Parameter Efficient Fine-Tuning (PEFT) methods enhance model adaptation by updating only a subset of the model's parameters, contrasting with full fine-tuning and transfer learning that either updates all parameters or makes minimal adjustments. Various PEFT approaches have been explored in the literature, such as Additive PEFT by \cite{houlsby2019parameter} introduced additional trainable components into a pre-trained model. Other approaches have been further explored to refine the concept of adapters with various designs and architecture, demonstrating their effectiveness across different tasks \cite{pfeiffer2020adapterfusion}, \cite{chen2024conv}, \cite{dhakal2024vlsm}.

\section{Methodology}
\subsection{Overview}
Given a multi-modal 3D volume $I \in \mathbb{R}^{C\times H \times W \times D}$ representing multiple MRI sequences, we want to obtain a segmentation mask $S \in \mathbb{N}^{N \times H \times W \times D}$ where N represents the number of segmentation classes. We assume availability of dataset $D = \{(I_i,S_i):i \in 1,2,\ldots,k\}$. In our case, $C=4$ representing 4 different MRI sequences i.e. T1-weighted (T1w), T1-weighted contrast-enhanced (T1-c), T2-weighted (T2w), T2w- Fluid-Attenuated Inversion Recovery (FLAIR) and $N=4$ representing segmentation classes i.e. background, Enhancing Tumor(ET), Non-Enhancing Tumor Core(NETC) and Surrounding Non-Enhancing Flair Hyper-intensity(SNFH).
\subsection{Dataset}
The BraTS 2021 \cite{baid2021rsna} and BraTS Africa \cite{adewole2023brain} were used in this study. BraTS 2021 dataset consists of 1251 pre-operative adult glioma cases. Similarly, the BraTS Africa dataset consists of 75 patients pre-operative adult glioma cases, collected retrospectively from various imaging centers in Africa\footnote{This data collection work was supported by Consortium for Advancement of MRI Education and Research in Africa (CAMERA) and funded by Lacuna Fund.}. Besides, the lower MRI spatial resolution, the BraTS Africa dataset is unique in its late presentation of disease and additional comorbidity resulting in a significant shift in distribution compared to similar datasets from western populations\cite{menze2014multimodal}.
All images are roughly aligned(co-registered) and resampled to isotropic voxels of $1mm^3$ as outlined by the BraTS challenge organizers \cite{baid2021rsna}\cite{adewole2023brain}. The images were segmented by trained experts with varying skill levels and then verified by board-certified radiologists with $>5$ years of experience. Detailed information on the dataset are available at \cite{adewole2023brain}.

\begin{figure}[ht]
    \centering
    \includegraphics[width=\textwidth]{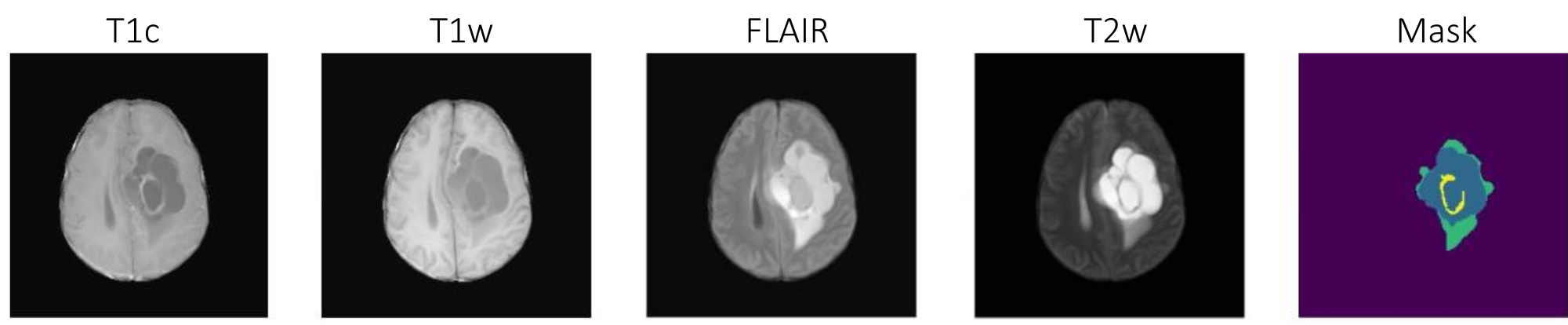}
    \caption{Brain image slices of a representative case from the BraTS Africa dataset with the four MRI modalities and manual annotated subregions (Mask), representing brain tumor sub-regions: Left to Right; T1-contrast-enhanced (T1c), pre-contrast T1-weighted (T1w), FLAIR, T2-weighted (T2w), and Mask}
    \label{fig:dataset-vis}
\end{figure}

\subsection{Preprocessing}
The dataset undergoes several spatial and intensity transformations to ensure consistency and improved model performance.
\begin{enumerate}
    \item Conversion to Canonical Orientation: Since the dataset combines two sets of MRI scans in various orientations, all images are transformed to a standardized Right-Anterior-Superior (RAS) orientation.
    % \item Resampling: The voxel spacing is resampled to achieve a uniform size of (1.0, 1.0, 1.0) across x, y, and z dimensions.
    \item Image Resizing: The images are resized to a target size of $128\times128\times128$ from its input volume of $255\times255\times255$, maintaining a consistent input size for the segmentation model. TorchIO's resize was used to downsample to the target size to reduce the image resolution through interpolation rather than cropping.
    \item Intensity Normalization: The intensity values are normalized using z normalization, which sets mean to zero and the standard deviation to one.
    \item Data Augmentation: Different augmentation techniques such as flipping along Left-Right axis, affine transformation (scaling, rotation, translation) and RandomNoise were applied.
\end{enumerate}

\subsection{Model Architecture} \label{Model Architecture}
We implement MedNeXt-S, a small-size MedNext architecture, a light-weight ConvNeXt\cite{liu2022convnet} architecture, as a baseline which implements encoder and decoder mechanism like in Unet\cite{ronneberger2015u}. But, unlike Unet, MedNeXt uses basic blocks that use modern design choices (such as inverted bottleneck, and transformer-inspired configurations) that improve over Unet. The Encoder consists of sequential multiple basic blocks that generate correlated spatial information with down-sampling to encode the input into rich feature maps by reducing spatial resolution, expanding feature maps, and the decoder combines multiple basic blocks with up-sampling to reconstruct segmentation from encoded feature maps. Each basic block consists of depth-wise convolution followed by expansion and reduction of channels forming an inverted bottleneck design with additional residual connection. Skip connection is implemented between consecutive spatial resolution encoder and up-sampling to restore lost spatial information during down-sampling.

In depth-wise convolution, richer feature maps are learned from each channel separately (without mixing across-channel features), and then these individual channel-wise features are combined using a parameter-efficient 1x1x1 convolution. Then, group normalization\cite{wu2018group} is applied to reduce the change in the distribution of inputs for stable training even in small batch\cite{roy20222} size. The depth-wise convolution is followed by the expansion of the channel dimension of the feature maps to obtain hierarchically connected expanded features required for downstream task. This is followed by GELU\cite{hendrycks2016gaussian} activation function to enhance convergence, and compression layer to obtain refined feature maps preserving only the important features in this layer. Additionally, residual connection is added to recover better gradient. 

\begin{figure}[!ht]
    \centering
    \includegraphics[width=0.70\textwidth]{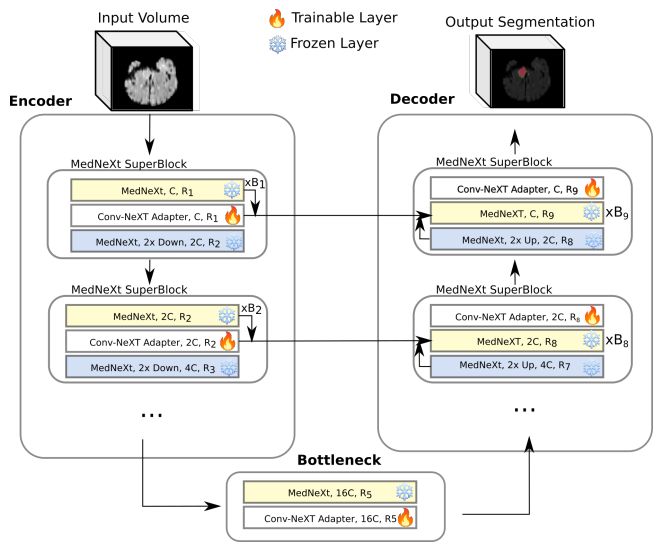}
    \caption{Conv-Adapter Inspired MedNext Segmentation Model Adapter Architecture}
    \label{fig:peft-mednext-label}
\end{figure}

We implemented MedNeXt-S with  kernel size of 3x3x3 and 32 input channels on input size 128x128x128. We applied expansion ratio 2 in each MedNeXt blocks(basic, up-sampling and down-sampling blocks). Similar to baseline architecture, applied two sequential basic blocks (B1 - B9, as in Figure~\ref{fig:peft-mednext-label}) 

% Baseline configure four different models MedNeXt-S, MedNeXt-B, MedNeXt-M and MedNeXt-L.\cite{roy2023mednext} The basic model, MedNeXt-S uses 32 channels, kernel size 3x3x3 with block count (repeated layer) and expansion ratio 2. Progressively large model B, M and L made some changes on expansion ratio and block count. MedNeXt-B varying expansion ratio, MedNeXt-M changes block count and MedNeXt-L update expansion ratio and block count with respect to its smaller version. Author also tested the MedNeXt versions using kernel:\{3,5\}\cite{roy2023mednext}, on S and B version kernel 5 perform better but not in M and L version of MedNeXt.
% We are implementing MedNeXt-S version with 3x3x3 kernel on input size 128x128x128.

\subsection{Proposed Architecture with Adapter}
Our proposed architecture integrates the MedNeXt backbone with PEFT using convolutional adapter blocks. Let \( f \) represent the intermediate representation learned by our pre-trained model, which are kept frozen, and \( f' \) be the adapted features or output from our trainable adapter module. The adapter features can be expressed as: 
\begin{equation*}
    f' = f + \text{Adapter}(f)
\end{equation*}
where `\( + \)' signifies the skip connection.

Previous works on PEFT modules by \cite{dhakal2024vlsm} used two linear modules in their architecture. The first module projects the original \( d \)-dimensional features into \( d' \) dimensions. The second linear module then projects \( d' \) back to \( d\)-dimensions, represented by:
\begin{equation*}
f' = f + \sigma( \psi(f \cdot W1) \cdot W2)    
\end{equation*}
where \( W1 \in \mathbb{R}^{d \times d'} \), \( W2 \in \mathbb{R}^{d' \times d} \), \( d' \leq d \), and \( \sigma \),\( \psi \) are some activation functions.

Similarly, the conv-adapter proposed by \cite{chen2024conv} introduces depthwise separable convolutions as an adapter. These convolutions decompose standard convolutions into depthwise and point-wise convolutions, significantly reducing the number of parameters and computations. The formulation is given by:
\begin{equation*}
f' = f + W_{\text{up}} \otimes \text{activation}(W_{\text{down}} \mathbin{\hat{\otimes}} f) = PL \left( \text{activation} \left( DC(f) \right) \right)   
\end{equation*}
where \( W_{\text{up}} \) and \( W_{\text{down}} \) are the convolutional weights. Here, \(\otimes\) denotes the depth-wise convolution, and \(\mathbin{\hat{\otimes}}\) denotes the point-wise convolution.

Our proposed architecture has an expansion layer between the depthwise and point-wise convolutions, inspired by the MedNeXt block. This adapted features with this expansion layer is represented as:
\begin{equation*}
f' = PL \left( \text{GELU} \left( EL \left( \text{LN} \left( DC(f) \right) \right) \right) \right)   
\end{equation*}
where \( PL \) denotes the projection layer that performs point-wise convolution, \( EL \) represents the expansion layer, \( \text{LN} \) is Layer Norm, GELU is activation applied and \( DC \) is Depthwise Convolution.

\begin{figure}
  \centering
  % \begin{subcaption}
    \centering
    \parbox[b]{.3\textwidth}{%
    \centering
    \includegraphics[width=0.3\textwidth]{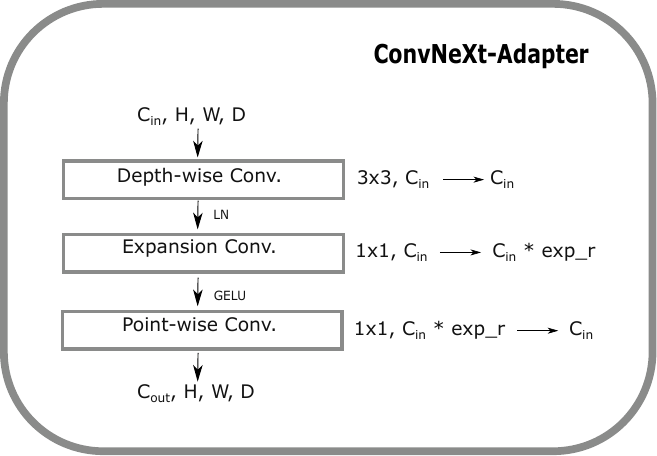}
    % \caption{ConvNeXt-Adapter}\label{fig:adapter-detail}
    }%
    \parbox[b]{.7\textwidth}{%
    \centering
    \includegraphics[width=0.6\textwidth]{images/conv-adapter-inspired-MedNeXT-PEFT-arch-types.pdf}
    % \caption{Conv-Adapter Inspired MedNext Segmentation Model Adapter Architecture}\label{fig:peft-mednext-label}
    }%
  % \end{subcaption}
  \caption{\textit{Left}: ConvNeXt-Adapter. \textit{Right}: Different forms of Adapter placement within the MedNeXT Superblock}
  \label{fig:arch-details}
\end{figure}

% \begin{figure}[b]
%     \centering
%     \includegraphics[width=0.4\textwidth]{images/convnext_adapter_detail.pdf}
%     \caption{ConvNeXt-Adapter}
%     \label{fig:adapter-detail}
% \end{figure}

\vspace{-20pt}
We experimented with two variations of adapter placement: Sequential and Parallel, as shown in figure~\ref{fig:arch-details}. However, Parallel placement of adapter resulted in lower performance (Average Dice of 0.78 as compared to 0.80 of Sequential Adapter). Therefore, the metrics mentioned in the Results and Discussion section are from the Sequential adapter. For reference, we have also provided a table comparing both versions in the Supplementary Materials.

The adapter layer results only in an additional 11.2\% increase in \#parameters (total 34.99M) which is still far more lightweight compared to parameter-heavy architectures like SwinTransformerV2($\sim$3B).  Additionally, finetuning adapter layer required $\sim$4 hrs compared to $\sim$10 hrs for full-finetuning.           

% \begin{figure}[H]
%     \centering
%     % \begin{subfigure}
%     \includegraphics[width=0.65\linewidth]{images/conv-adapter-inspired-MedNeXT-PEFT-arch-types.pdf}
%     \caption{Different forms of Adapter placement within the MedNeXT Superblock}
%     \label{fig:adapt-placem-form}
%     % \end{subfigure}

% \end{figure}

\section{Results and Discussion}
As shown in Table~\ref{table:evaluation-on-2021_vs_ssa}, the MedNeXt model trained on BraTS-2021 performs comparatively well on in-domain evaluation set whereas when evaluated on a samples from different population, BraTS-Africa in this case, unseen during training, we see drastic reduction in performance. This necessitates further intervention for the model to be usable on the downstream BraTS-Africa task. We explore Parameter-efficient Fine-tuning (PEFT) for leveraging BraTS-2021 dataset due to its proximity to the downstream task and added computational efficiency. The Lesionwise metric differs from vanilla Dice score in that it comparatively heavily penalizes absence of individual lesion. The evaluation was performed by MedPerf, a standardized platform ensuring reproducible benchmarking on site’s dataset\cite{karargyris2023federated}

\begin{table}[!ht]
    \centering
    \small
    \caption{Avg. Validation Dice Score on BraTS-2021 and BraTS-Africa validation dataset when trained on BraTS-2021}
    \label{table:evaluation-on-2021_vs_ssa}
    \begin{tabular}{|c|c|c|c|}
        \hline
         & BraTS-2021 & BraTS-Africa\\ \hline
        Average Dice & 0.70 & 0.50\\ \hline    
    \end{tabular}
\end{table}

As shown in Table~\ref{table:with_vs_without_peft}, the proposed Conv-adapter-based finetuning results in superior segmentation performance compared to the baseline MedNeXt model trained only on the BraTS-Africa dataset.  The Conv-adapter-based fine-tuned model is obtained by first pre-training the MedNeXt architecture on the BraTS-2021 dataset and then finetuned on BraTS-Africa. Both baseline and fine-tuned models are evaluated on the BraTS-Africa validation dataset. The statistical significance of the observed values was evaluated, yielding P-value of 0.000116, confirming that the differences are highly significant between performance of “without Fine-Tuning” vs “with PEFT Fine-Tuning”.

% \begin{table}[H]
%     \centering
%     \small
%     \caption{Performance comparison without fine-tuning and with PEFT fine-tuning.}
%     \label{table:with_vs_without_peft}
%     \begin{tabular}{|c|c|c|c|c|c|c|c|c|c|}
%         \hline
%         \multirow{2}{*}{Network} & \multirow{2}{*}{Metric} & \multicolumn{4}{c|}{Without Fine-Tuning} & \multicolumn{4}{c|}{With PEFT Fine-Tuning} \\ \cline{3-10}
%         & & ET & TC & WT & Avg. & ET & TC & WT & Avg. \\ \hline
%         \multirow{3}{*}{\begin{tabular}[c]{@{}c@{}}MedNeXt-S\end{tabular}}
%         & LesionWise Dice \, $\uparrow$ & 0.44 & 0.35 & 0.26 & 0.35 & \textbf{0.65} & \textbf{0.68} & \textbf{0.81} & \textbf{0.71} \\ \cline{2-10}
%         & Dice \, $\uparrow$ & 0.72 & 0.68 & 0.77 & 0.72 & \textbf{0.74} & \textbf{0.79} & \textbf{0.88} & \textbf{0.80} \\ \cline{2-10}
%         & LesionWise HD95 $\downarrow$ & 167.76 & 201.69 & 255.18 & 208.21 & \textbf{74.40} & \textbf{79.95} & \textbf{34.31} & \textbf{62.89} \\ \cline{2-10}
%         & HD95 $\downarrow$ & 33.98 & 36.74 & 33.49 & 34.73 & \textbf{28.84} & \textbf{29.62} & \textbf{7.15} & \textbf{21.87} \\ \hline       
%     \end{tabular}
% \end{table}

\begin{table}
    \centering
    \small
    \caption{Performance comparison between without fine-tuning and with PEFT fine-tuning.}
    \label{table:with_vs_without_peft}
    \resizebox{\textwidth}{!}{
    \begin{tabular}{|c|c|c|c|c|c|c|c|c|c|}
        \hline
        \multirow{2}{*}{Network} & \multirow{2}{*}{Metric} & \multicolumn{4}{c|}{Without Fine-Tuning} & \multicolumn{4}{c|}{With PEFT Fine-Tuning} \\ \cline{3-10}
         &  & ET & TC & WT & Avg. & ET & TC & WT & Avg. \\ \hline
        \multirow{7}{*}{MedNeXt-S} 
        & LesionWise Dice \, $\uparrow$ & 0.44 & 0.35 & 0.26 & 0.35 & \textbf{0.65} & \textbf{0.68} & \textbf{0.81} & \textbf{0.71} \\ 
        & \textit{Std. Dev.} & (0.29) & (0.25) & (0.20) & (0.25) & (0.27) & (0.30) & (0.22) & (0.26) \\ \cline{2-10}
        & Dice \, $\uparrow$ & 0.72 & 0.68 & 0.77 & 0.72 & \textbf{0.74} & \textbf{0.79} & \textbf{0.88} & \textbf{0.80} \\ 
        & \textit{Std. Dev.} & (0.21) & (0.23) & (0.22) & (0.22) & (0.21) & (0.23) & (0.14) & (0.19) \\ \cline{2-10}
        & LesionWise HD95 \, $\downarrow$ & 167.76 & 201.69 & 255.18 & 208.21 & \textbf{74.40} & \textbf{79.95} & \textbf{34.31} & \textbf{62.89} \\ 
        & \textit{Std. Dev.} & (128.50) & (110.80) & (87.81) & (109.04) & (117.04) & (118.13) & (73.98) & (103.05) \\ \cline{2-10}
        & HD95 \, $\downarrow$ & 33.98 & 36.74 & 33.49 & 34.73 & \textbf{28.84} & \textbf{29.62} & \textbf{7.15} & \textbf{21.87} \\ 
        & \textit{Std. Dev.} & (65.17) & (63.51) & (22.67) & (50.45) & (87.01) & (86.50) & (7.48) & (60.33) \\ \hline       
    \end{tabular}
    }
\end{table}

As shown in Table~\ref{table:full-ft_vs_peft}, the proposed PEFT-based fine-tuned model slightly surpasses the full-fine tuned model, on average. We hypothesize that this is due to the small size of the fine-tuned dataset which results in a parameter-efficient method to prevent overfitting, a lingering challenge for full-fine tuned methods. On the other hand, Lesionwise Dice for smaller sub-regions such as enhancing tumor or non-enhancing tumor core are better when full-finetuned, whereas PEFT excels at whole tumor which is a comparatively larger structure. Overall, the P-value of 0.63 indicates that the observed difference between “Full Fine-Tuning” and “with PEFT Fine-Tuning” is not statistically significant, suggesting comparable performance of the two methods.

% \begin{table}[H]
%     \centering
%     \small
%     \caption{Performance comparison between Full Fine-Tuning and PEFT Fine-Tuning.}
%     \label{table:full-ft_vs_peft}
%     \begin{tabular}{|c|c|c|c|c|c|c|c|c|c|}
%         \hline
%         \multirow{2}{*}{Network} & \multirow{2}{*}{Metric} & \multicolumn{4}{c|}{Full Fine-Tuning} & \multicolumn{4}{c|}{PEFT Fine-Tuning} \\ \cline{3-10}
%         & & ET & TC & WT & Avg. & ET & TC & WT & Avg. \\ \hline
%         \multirow{3}{*}{\begin{tabular}[c]{@{}c@{}}MedNeXt-S\end{tabular}} 
%         & LesionWise Dice $\uparrow$ & \textbf{0.67} & \textbf{0.69} & 0.75 & 0.70 & 0.65 & 0.68 & \textbf{0.81} & \textbf{0.71} \\ \cline{2-10}
%         & Dice $\uparrow$ & 0.73 & 0.77 & 0.83 & 0.77 & \textbf{0.74} & \textbf{0.79} & \textbf{0.88} & \textbf{0.80} \\ \cline{2-10}
%         & LesionWise HD95 $\downarrow$ & \textbf{61.43} & \textbf{67.53} & 57.27 & \textbf{62.08} & 74.40 & 79.95 & \textbf{34.31} & 62.89 \\ \cline{2-10}
%         & HD95 $\downarrow$ & 30.12 & 31.73 & 20.76 & 27.54 & \textbf{28.84} & \textbf{29.62} & \textbf{7.15} & \textbf{21.87} \\ \hline         
%     \end{tabular}
% \end{table}

\begin{table}[!ht]
    \centering
    \small
    \caption{Performance comparison between Full Fine-Tuning and with PEFT Fine-Tuning.}
    \label{table:full-ft_vs_peft}
    \resizebox{\textwidth}{!}{
    \begin{tabular}{|c|c|c|c|c|c|c|c|c|c|}
        \hline
        \multirow{2}{*}{Network} & \multirow{2}{*}{Metric} & \multicolumn{4}{c|}{Full Fine-Tuning} & \multicolumn{4}{c|}{With PEFT Fine-Tuning} \\ \cline{3-10}
        &  & ET & TC & WT & Avg. & ET & TC & WT & Avg. \\ \hline
        \multirow{7}{*}{MedNeXt-S} 
        & LesionWise Dice \, $\uparrow$ & \textbf{0.67} & \textbf{0.69} & 0.75 & 0.70 & 0.65 & 0.68 & \textbf{0.81} & \textbf{0.71} \\ 
        & \textit{Std. Dev.} & (0.26) & (0.28) & (0.25) & (0.26) & (0.27) & (0.30) & (0.22) & (0.26) \\ \cline{2-10}
        & Dice \, $\uparrow$ & 0.73 & 0.77 & 0.83 & 0.77 & \textbf{0.74} & \textbf{0.79} & \textbf{0.88} & \textbf{0.80} \\ 
        & \textit{Std. Dev.} & (0.21) & (0.24) & (0.17) & (0.21) & (0.21) & (0.23) & (0.14) & (0.19) \\ \cline{2-10}
        & LesionWise HD95 \, $\downarrow$ & \textbf{61.43} & \textbf{67.53} & 57.27 & \textbf{62.08} & 74.40 & 79.95 & \textbf{34.31} & 62.89 \\ 
        & \textit{Std. Dev.} & (111) & (112.26) & (99.54) & (107.60) & (117.04) & (118.13) & (73.98) & (103.05) \\ \cline{2-10}
        & HD95 \, $\downarrow$ & 30.12 & 31.73 & 20.76 & 27.54 & \textbf{28.84} & \textbf{29.62} & \textbf{7.15} & \textbf{21.87} \\ 
        & \textit{Std. Dev.} & (87.22) & (86.79) & (63.12) & (79.04) & (87.01) & (86.50) & (7.48) & (60.33) \\ \hline         
    \end{tabular}
    }
\end{table}

As shown in Table~\ref{table:performance_comparison}, although full fine-tuning resulted in more consistent performance, PEFT achieved higher average performance while maintaining high specificity (0.99) but lower sensitivity (0.75).

% \begin{table}[H]
%     \centering
%     \small
%     \caption{Sensitivity and specificity of BraTS-SSA without Fine-Tuning, with Full-Finetuning, and with PEFT fine-tuning}
%     \label{table:performance_comparison}
%     \begin{tabular}{|c|c|c|c|c|c|c|c|c|c|}
%         \hline
%         & \multicolumn{3}{c|}{Without Fine-Tuning} & \multicolumn{3}{c|}{With Full Fine-Tuning} & \multicolumn{3}{c|}{With PEFT Fine-Tuning} \\ \hline
%         & ET & TC & WT & ET & TC & WT & ET & TC & WT \\ \hline
%         Sensitivity $\uparrow$ & 0.40 & 0.27 & 0.36 & 0.64 & 0.59 & 0.68 & \textbf{0.67} & \textbf{0.66} & \textbf{0.75} \\ \hline   
%         Specificity $\uparrow$ & 0.99 & 0.98 & 0.96 & 0.99 & 0.99 & 0.99 & \textbf{0.99} & \textbf{0.99} & \textbf{0.99} \\ \hline
%     \end{tabular}
% \end{table}

\begin{table}[!ht]
    \centering
    \small
    \caption{Sensitivity and Specificity of BraTS-Africa Without Fine-Tuning, With Full Fine-Tuning, and With PEFT Fine-Tuning}
    \label{table:performance_comparison}
    \resizebox{\textwidth}{!}{
    \begin{tabular}{|c|c|c|c|c|c|c|c|c|c|}
        \hline
        & \multicolumn{3}{c|}{Without Fine-Tuning} & \multicolumn{3}{c|}{With Full Fine-Tuning} & \multicolumn{3}{c|}{With PEFT Fine-Tuning} \\ \hline
        & ET & TC & WT & ET & TC & WT & ET & TC & WT \\ \hline
        Sensitivity $\uparrow$ & 
        0.40 & 0.27 & 0.36 & 0.64 & 0.59 & 0.68 & \textbf{0.67} & \textbf{0.66} & \textbf{0.75} \\ 
        \textit{Std. Dev.} & 
        (0.37) & (0.34) & (0.41) & (0.31) & (0.38) & (0.38) & \textbf(0.27) & \textbf(0.33) & \textbf(0.35) \\ \hline   
        Specificity $\uparrow$ & 
        0.99 & 0.98 & 0.96 & \textbf{0.99} & \textbf{0.99} & \textbf{0.99} & \textbf{0.99} & \textbf{0.99} & \textbf{0.99} \\ 
        \textit{Std. Dev.} & 
        (0.01) & (0.03) & (0.06) & \textbf(0.002) & \textbf(0.004) & (0.01) & (0.001) & \textbf(0.004) & \textbf(0.007) \\ \hline
    \end{tabular}
    }
\end{table}

\begin{figure}[!ht]
    \centering
    % \subfloat[\centering label 1]{{\includegraphics[width=0.45\textwidth]{images/average_dice.pdf} }}%
    % \qquad
    {\includegraphics[width=0.5\textwidth]{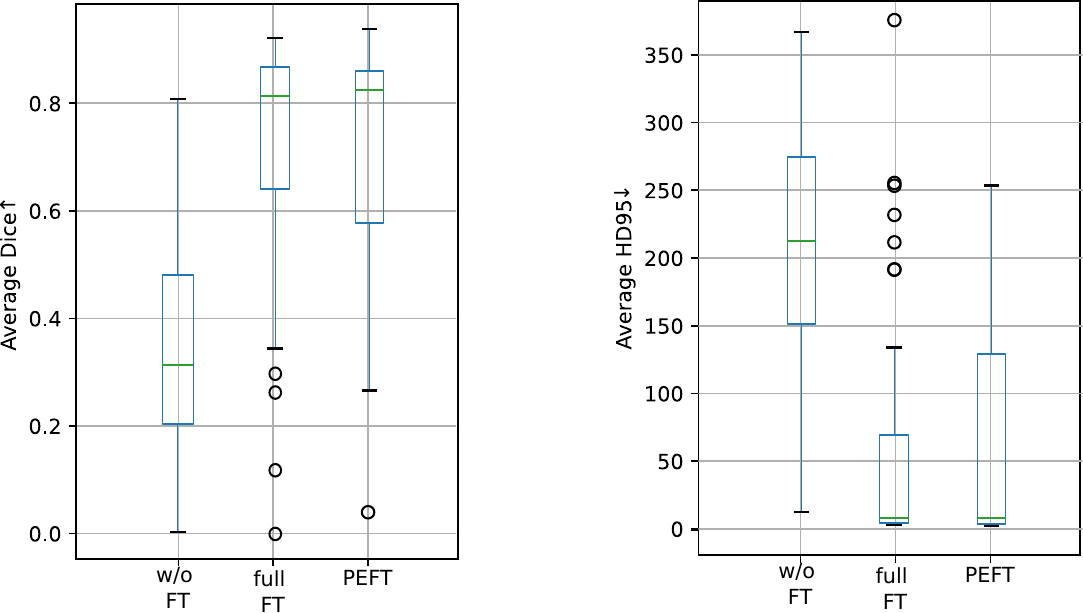} }%
    \caption{Boxplot comparison of segmentation methods: without fine-tuning, full fine-tuning, and PEFT, using Average Dice (left) and Average HD95 (right)}
    \label{fig:boxplot-label}
\end{figure}

% \paragraph{Visual Comparison}

% \begin{figure}[h]
%     \centering
%     \includegraphics[width=0.65\textwidth]{images/visual_comparison_color_coded-v2.pdf}
%     \caption{Visual comparison shows Parameter-efficient fine-tuned segmentation considerably improves performance compared to training only on the downstream (BraTS Africa) dataset.}
%     \label{fig:results-vis}
% \end{figure}
\begin{figure}[!ht]
    \centering
    \includegraphics[width=0.95\textwidth]{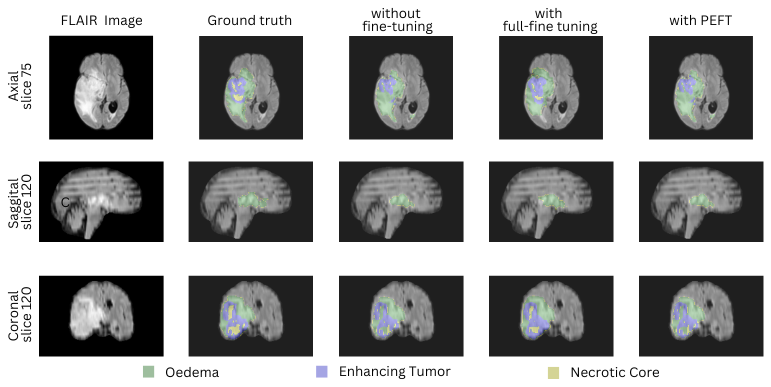}
    \caption{Visual comparison of Predicted and Ground Segmentations with and without finetuning.}
    \label{fig:results-vis}
\end{figure}
\section{Conclusion}
We explored convolutional adapter layer for parameter efficient fine tuning in light-weight MedNeXt-S architecture. The proposed method showed that PEFT can leverage both BraTS-2021 and BraTS Africa dataset to obtain performance comparable to Full Fine-tuning with reduced training time.
\newline
\newline
\newline
\section{Acknowledgement}
The authors thank Sprint AI Training for African Medical Imaging Knowledge Translation (SPARK) Academy in Deep Learning and Medical Imaging 2024 Team for helping build foundational knowledge on Medical Image Segmentation including clinical domain-specific lecture sessions that helped us appreciate the complexity of the problem. The compute facility was provided by Digital Research Alliance of Canada (Compute Canada). Additional compute support was provided by Nepal Applied Mathematics and Informatics Institute for research (NAAMII). Finally, we would like to thank Lacuna Fund for Health and Equity, the Radiological Society of North America (RSNA), Research \& Education (R\&E) Foundation Derek Harwood-Nash International Education Scholar Grant, and National Science and Engineering Research Council of Canada (NSERC) Discovery Launch Supplement for making the SPARK Academy possible via research grant supports.
\bibliographystyle{splncs04}
\bibliography{main}
\clearpage
\section{Supplementary Materials}
\subsection{Performance of Parallel Vs Sequential ConvNeXt Adapter}

\begin{table}
    \centering
    \small
    \caption{Performance comparison between PEFT using Sequential ConvNeXt Adapter and Parallel ConvNeXt Adapter.}
    \label{table:seq-vs-parallel}
    \begin{tabular}{|c|c|c|c|c|c|c|c|c|c|}
        \hline
        \multirow{2}{*}{Network} & \multirow{2}{*}{Metric} & \multicolumn{4}{c|}{Parallel ConvNeXt Adapter} & \multicolumn{4}{c|}{Sequential ConvNeXt Adapter} \\ \cline{3-10}
        & & ET & TC & WT & Avg. & ET & TC & WT & Avg. \\ \hline
        \multirow{3}{*}{MedNeXt-S} 
         & LesionWise Dice \, $\uparrow$ & 0.59 & 0.61 & 0.54 & 0.58 & 0.65& 0.68& 0.81& 0.71\\ \cline{2-10}
        & Dice \, $\uparrow$ & 0.72 & 0.77 & 0.86 & 0.78 & 0.74& 0.79& 0.88& 0.80\\ \cline{2-10}
        & LesionWise HD95 $\downarrow$ & 90.44 & 96.96 & 146.19 & 111.20 & 74.40& 79.95& 34.31& 62.89\\ \cline{2-10}
        & HD95 $\downarrow$ & 29.96 & 32.52 & 24.41 & 28.96 & 28.84& 29.62& 7.15 & 21.87 \\ \hline        
    \end{tabular}
\end{table}

\subsection{Qualitative Visualization}
\begin{figure}
    \centering
    \includegraphics[width=\textwidth]{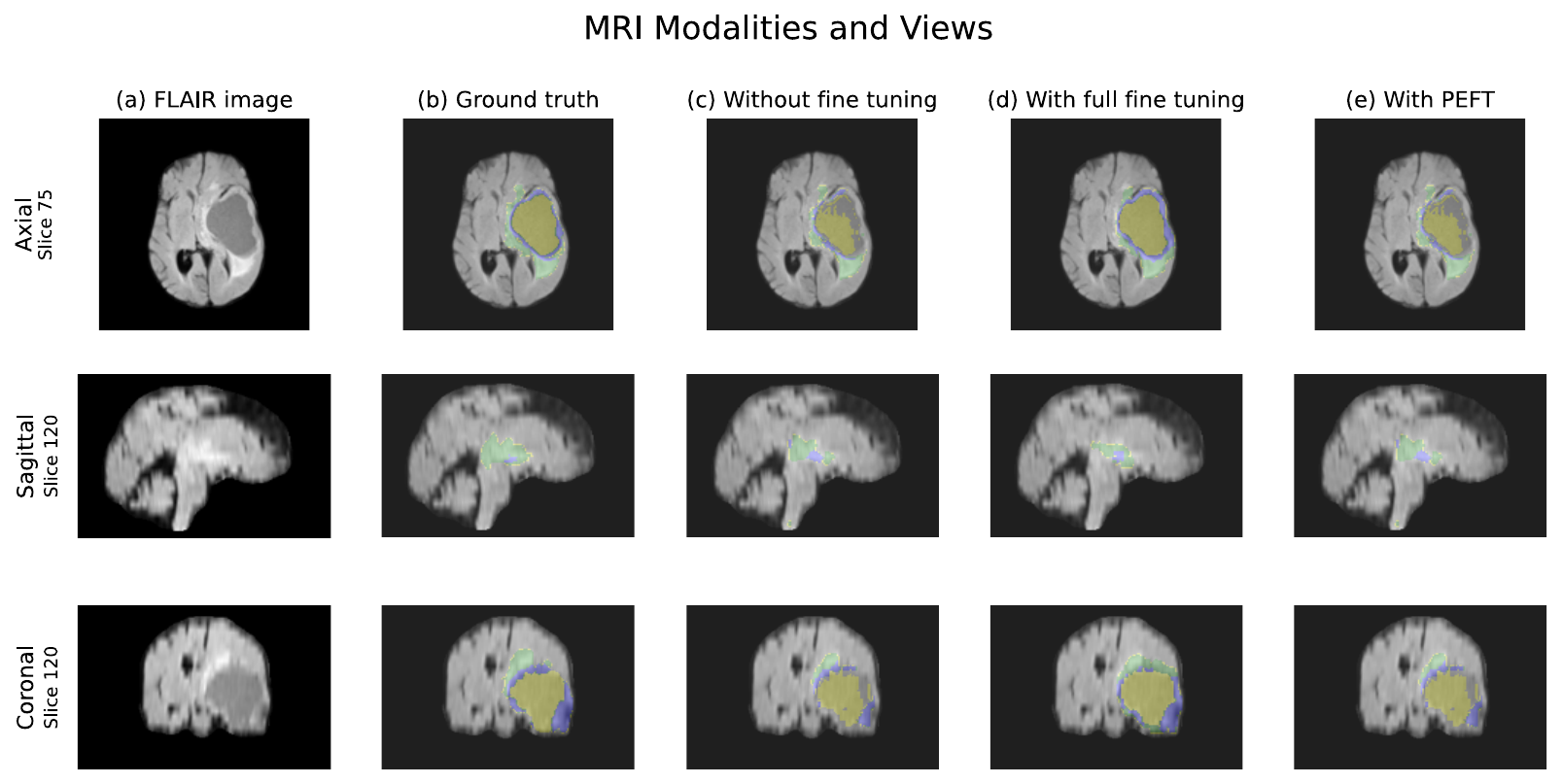}
    \caption{Visual comparison of Predicted and Ground Segmentations with and without finetuning}
\end{figure}
% \begin{figure}[H]
%     \centering
%     \includesvg[width=0.95\linewidth, height=0.4\textheight]{images/vimages/00129-000.svg}
%     \caption{Visual comparison of Predicted and Ground Segmentations with and without finetuning. Left to Right: FLAIR Image, Ground Segmentation Mask, Predicted segmentation mask by model trained on BraTS Africa dataset only, and Predicted segmentation mask with PEFT}
% \end{figure}
% \begin{figure}[H]
%     \centering
%     \includesvg[width=0.95\linewidth, height=0.4\textheight]{images/vimages/00132-000.svg}
%     \caption{Visual comparison of Predicted and Ground Segmentations with and without finetuning}
% \end{figure}

\begin{figure}
    \centering
    \includegraphics[width=\textwidth]{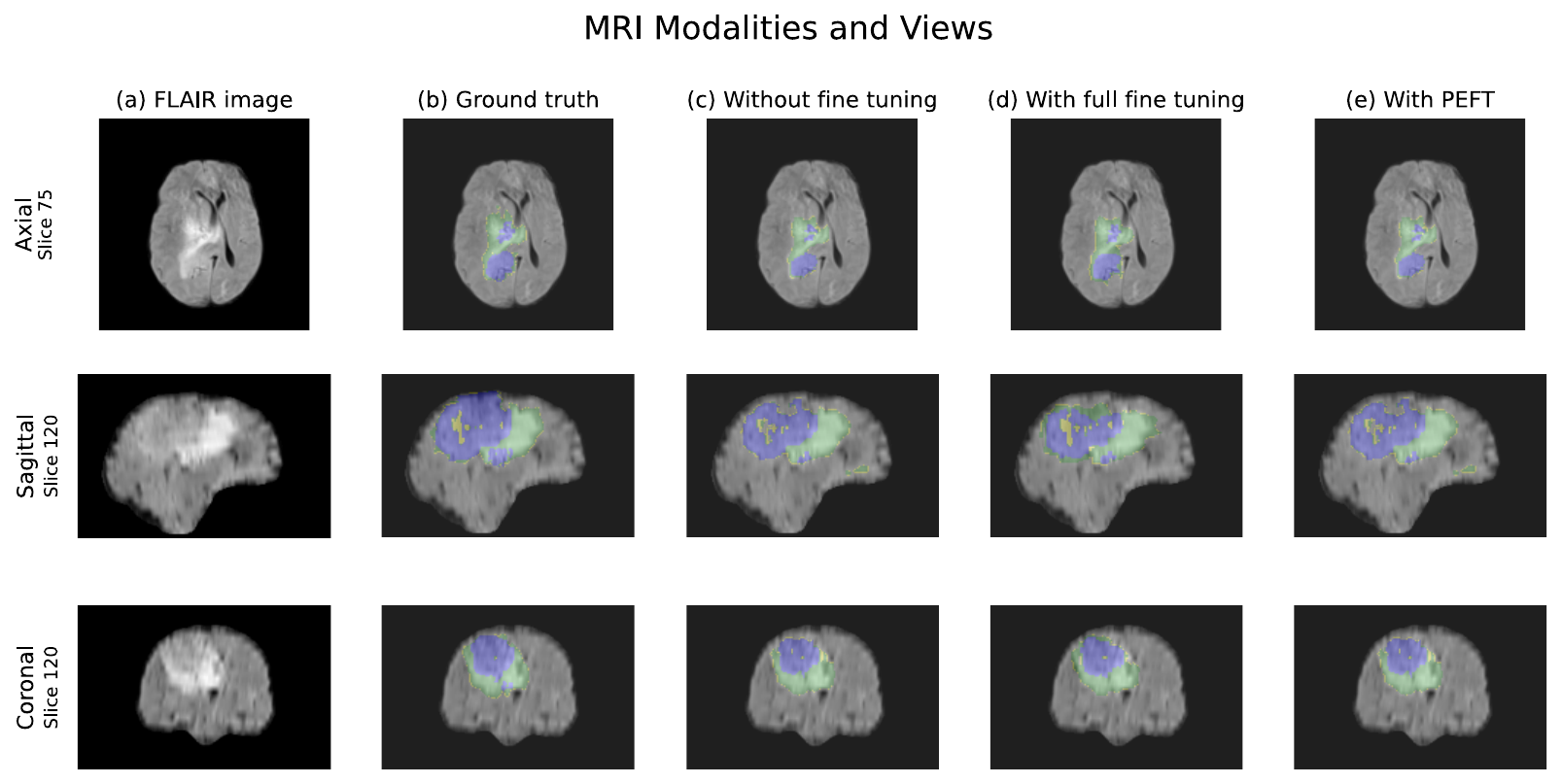}
    \caption{Visual comparison of Predicted and Ground Segmentations with and without finetuning}
\end{figure}
\end{document}